\RequirePackage{fix-cm}
\documentclass{externalized}
\usepackage{microtype}
\usepackage{graphicx}
\usepackage{hyperref}
\usepackage{url}
\usepackage{booktabs}
\usepackage{amsfonts}
\usepackage{amsmath}
\usepackage{amssymb}
\usepackage{amsthm}
\usepackage{mathtools}
\usepackage{enumitem}
\usepackage{multirow}
\usepackage{subcaption}
\usepackage{xcolor}
\usepackage{listings}
\usepackage{natbib}
\usepackage{multicol}
\usepackage{pifont}
\usepackage{xspace}
\usepackage{wrapfig}
\usepackage{float}
\usepackage{tabularx}
\usepackage{ragged2e}
\usepackage{longtable}
\usepackage{colortbl}

\DeclareTextFontCommand{\tabletextbf}{\fontseries{bx}\selectfont}
\AtBeginEnvironment{tabular}{\let\textbf\tabletextbf}
\AtBeginEnvironment{tabular*}{\let\textbf\tabletextbf}
\AtBeginEnvironment{tabularx}{\let\textbf\tabletextbf}
\AtBeginEnvironment{longtable}{\let\textbf\tabletextbf}

\providecommand{\noopsort}[1]{}
\AtBeginDocument{%
  \providecommand\BibTeX{{%
    \normalfont B\kern-0.5em{\scshape i\kern-0.25em b}\kern-0.8em\TeX}}}

\makeatletter
\DeclareRobustCommand\onedot{\futurelet\@let@token\@onedot}
\def\@onedot{\ifx\@let@token.\else.\null\fi}

\makeatother

\definecolor{codegreen}{rgb}{0,0.6,0}
\definecolor{codegray}{rgb}{0.5,0.5,0.5}
\definecolor{codepurple}{rgb}{0.58,0,0.82}
\definecolor{backcolour}{rgb}{0.95,0.95,0.92}

\lstdefinestyle{mystyle}{
  backgroundcolor=\color{backcolour},
  commentstyle=\color{codegreen},
  keywordstyle=\color{magenta},
  numberstyle=\tiny\color{codegray},
  stringstyle=\color{codepurple},
  basicstyle=\ttfamily\footnotesize,
  breakatwhitespace=false,
  breaklines=true,
  captionpos=b,
  keepspaces=true,
  numbers=left,
  numbersep=5pt,
  showspaces=false,
  showstringspaces=false,
  showtabs=false,
  tabsize=2
}

\hypersetup{%
  pdftitle={From Delegation to Coordination: Learning Efficient Multi-Agent Deep Research},
  pdfauthor={Yuyang Wu, Yang Liu, Zequn Xie, Junjie Wang, Dan Yang, Zhehao Tan, Ding-Chu Zhang, Junming Yang, Yue Shen, Jian Wang, Jinjie Gu},
  pdfkeywords={multi-agent systems, deep research, reinforcement learning, credit assignment, LLM agents},
}

\newcommand{\YangLiu}[1]{#1}

\newcommand{\logorule}{%
  \raisebox{\dimexpr-0.5\height+1.0mm\relax}%
    {\textcolor{black!25}{\rule{0.5pt}{7.8mm}}}%
}
\renewcommand{\institutionlogosleft}{%
  \raisebox{\dimexpr-0.5\height+1.2mm\relax}{\includegraphics[height=7.5mm]{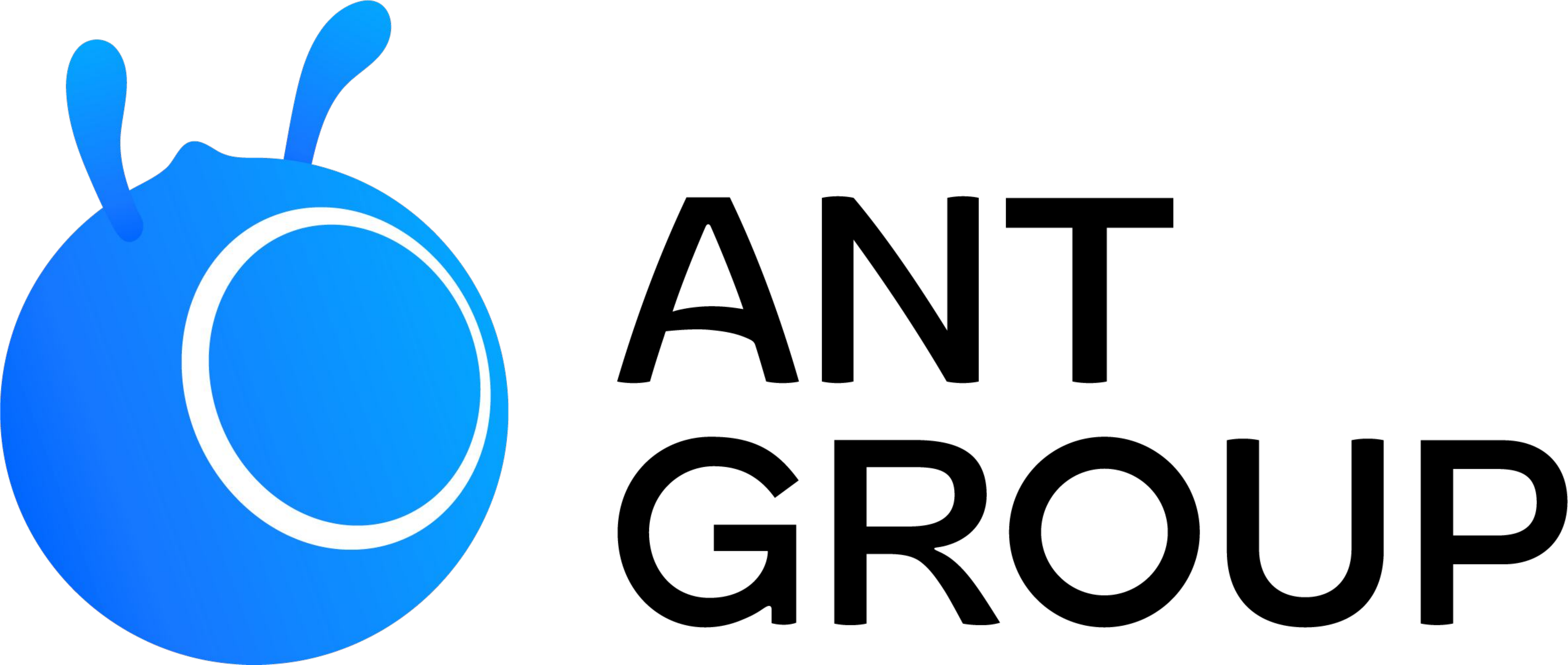}}%
  \hspace{4.5mm}\logorule\hspace{4.5mm}%
  \raisebox{\dimexpr-0.5\height+1.2mm\relax}{\includegraphics[height=7.5mm]{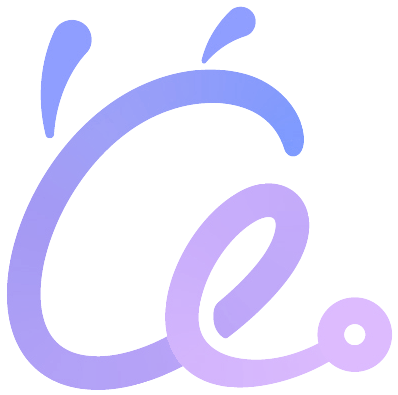}}%
}
\renewcommand{\institutionlogosright}{}

\title{From Delegation to Coordination: Learning Efficient Multi-Agent Deep Research}

\author[1,*]{Yuyang Wu}
\author[2,*]{Yang Liu}
\author[3,*]{Zequn Xie}
\author[1]{Junjie Wang}
\author[1]{Dan Yang}
\author[1]{Zhehao Tan}
\author[1]{Ding-Chu Zhang}
\author[4]{Junming Yang}
\author[1]{Yue Shen}
\author[1]{Jian Wang}
\author[1,\dagger]{Jinjie Gu}

\renewcommand{\affiliationlist}{%
  {\affiliationfont\affiliationfamily
    $^{1}$Ant Group, $^{2}$University of Electronic Science and Technology of China\par
    \vspace{0.6mm}%
    $^{3}$Zhejiang University, $^{4}$Southeast University\par}%
}

\contribution[*]{Equal contribution}
\contribution[\dagger]{Corresponding author}

\abstract{%
Multi-agent deep-research systems use delegation to parallelize search and
keep long tool traces out of the main agent's context. Yet delegation helps
only when task allocation and subtask execution are coordinated: the main agent
must issue useful subtasks, while subagents must return reliable evidence.
Existing training either optimizes only the main agent or broadcasts the same
team advantage to all agents, which can reward misleading subagent reports
and penalize useful ones. We introduce \textbf{LEAD-RL} (\textbf{L}earning
\textbf{E}ffective \textbf{A}gent \textbf{D}elegation with Reinforcement Learning),
a joint main-agent--subagent training framework with contribution-aware credit
assignment. A training-time rubric evaluates the research utility of each
subagent report and gates its inherited team advantage, while final-answer
correctness remains the only team reward; no token, step, or speedup reward is used.
Across five benchmarks and two model families, LEAD-RL improves answer quality.
On BrowseComp, it improves accuracy by 2.49 percentage points and reduces
critical-path cost by 24.5\% over joint training with shared team credit.
During validation, accuracy increases by 6.78 points at approximately constant
critical-path cost. These results show that contribution-aware credit assignment
improves the quality--efficiency trade-off of multi-agent deep research.%
}

\begin{document}
\raggedbottom
\maketitle

\setlength{\parskip}{4pt plus 1pt minus 1pt}

\section{Introduction}
\label{sec:intro}

Deep-research agents answer complex questions through iterative search,
tool use, verification, and cross-source synthesis
\citep{team2025tongyi,team2026mirothinker}. Their interaction histories
accumulate search results, intermediate hypotheses, and unsuccessful
attempts alongside useful evidence. Context-management methods compress
or selectively retain this growing state
\citep{wu2025resum,ye2025agentfold,lu2026longseeker}.
Delegation offers a complementary mechanism: \YangLiu{subagents} investigate
self-contained subtasks in isolated contexts and return condensed \YangLiu{subagent reports}.
Following prior swarm systems \citep{team2026kimi,ning2026searchswarm},
we study delegation as \emph{active context management}: the policy
controls where research is performed and what information returns to the
\YangLiu{main agent}'s global context.

Separate contexts do not guarantee effective delegation. The \YangLiu{main agent}
must scope useful subtasks and integrate their results; \YangLiu{subagents} must
search effectively and report reliable evidence. Poor coordination can
create redundant research, misleading \YangLiu{subagent reports}, or a long \YangLiu{subagent} branch
that delays the investigation. The learning problem is therefore to
improve task allocation and execution together, not simply to use more agents.

Existing approaches learn delegation from supervised trajectories
\citep{ning2026searchswarm}, train a \YangLiu{main agent} against frozen \YangLiu{subagents}
\citep{team2026kimi}, or jointly optimize multiple roles
\citep{mo2025matpo,xu2026wideseek}. \YangLiu{Joint main-agent--subagent training} enables co-adaptation
but leaves a credit-assignment ambiguity: successful rollouts can contain
misleading \YangLiu{subagent reports}, while unsuccessful rollouts can contain useful evidence
that the \YangLiu{main agent} fails to integrate. Broadcasting the same outcome
advantage to every \YangLiu{subagent} can thus reinforce harmful behavior or penalize
productive research.

\YangLiu{Our key observation is that delegation provides a natural granularity
for credit assignment: the subagent report. Each report corresponds to a
specific delegated brief and contains its own evidence and conclusion,
allowing its research utility to be assessed separately from the final team
outcome. Building on this structure, we introduce \textbf{LEAD-RL}
(\textbf{L}earning \textbf{E}ffective \textbf{A}gent \textbf{D}elegation with
Reinforcement Learning), which jointly updates a shared policy for the main
agent and subagents from complete swarm trajectories. A training-time rubric
classifies each subagent report as helpful, neutral, or harmful and gates its
inheritance of the team advantage. Helpful reports retain only positive team
advantages, harmful reports retain only negative ones, and neutral reports
retain the original advantage. The main agent continues to receive the team
advantage directly, so its delegation decisions remain supervised by final
task outcomes. The rubric therefore filters misassigned credit rather than
adding a process reward: final-answer correctness remains the sole team
objective, without explicit token, step, or speedup terms.}

We assess efficiency through critical-path cost, which compares execution
depth across systems under a common harness, and idealized speedup, which
measures parallelization of each trajectory's own workload rather than
wall-clock acceleration. Both are interpreted alongside answer quality
and total tokens, since parallelism alone need not reflect useful research.

Across five benchmarks and two model families, our trained models improve
answer quality and compete with substantially larger systems. Controlled
BrowseComp comparisons isolate the RL stage: LEAD-RL improves accuracy
by 2.49 percentage points and reduces critical-path cost by 24.5\%
relative to \YangLiu{joint main-agent--subagent training} with shared team credit. Validation accuracy also
improves at approximately constant critical-path cost. These results
suggest that distributing outcome credit more selectively can shape both
research quality and computational behavior without explicit cost rewards.

Our main contributions are:
\begin{itemize}[leftmargin=*]
    \item We develop a framework for delegation as \emph{active context
    management}, coupling task allocation, isolated \YangLiu{subagent} research,
    and the integration of condensed \YangLiu{subagent reports}.
    \item \YangLiu{We propose \emph{LEAD-RL}, combining joint
    main-agent--subagent training with report-level, rubric-based credit
    assignment under an outcome-only team reward.}
    \item We evaluate swarm efficiency through complementary absolute
    critical-path cost and relative idealized speedup, alongside answer
    quality and token consumption.
    \item We demonstrate gains across five benchmarks and two model
    families, with controlled RL comparisons and validation dynamics
    revealing a better accuracy--efficiency trade-off.
\end{itemize}

\section{Related Work}
\label{sec:related}

\subsection{Deep Research and Context Management}

\YangLiu{Deep-research agents are trained to acquire evidence through retrieval
and open-web interaction. Search-R1 and DeepResearcher establish this setting
\citep{jin2025searchr1,zheng2025deepresearcher}, while WebSailor, Tongyi
DeepResearch, REDSearcher, MiroThinker, and OpenSeeker extend its training data
and interaction strategies
\citep{li2025websailor,team2025tongyi,chu2026redsearcher,team2026mirothinker,du2026openseeker}.
Within this line, WebClipper uses graph-based pruning to remove redundant steps
from training trajectories \citep{wang-etal-2026-webclipper}, while SlimSearcher
combines trajectory filtering with adaptive reward gating to encourage agents
to use fewer tool calls and tokens \citep{xie-etal-2026-slimsearcher}.}

A complementary line addresses how evidence is retained as interaction
histories grow. ReSum summarizes state \citep{wu2025resum}, AgentFold consolidates
trajectories \citep{ye2025agentfold}, and LongSeeker provides context-editing
operations \citep{lu2026longseeker}. These methods transform a single history,
whereas Kimi K2.5 and SearchSwarm distribute research across isolated subagent
contexts \citep{team2026kimi,ning2026searchswarm}. We study contribution-aware
joint RL for this distributed form of active context management.

\subsection{Multi-Agent Delegation and Credit Assignment}

A main agent's task assignments determine what its subagents
investigate, while their reports shape its subsequent decisions.
SearchSwarm learns delegation from supervised trajectories
\citep{ning2026searchswarm}.
Kimi K2.5 trains the main agent with frozen subagents
\citep{team2026kimi}, whereas MATPO and WideSeek-R1 jointly
optimize both roles \citep{mo2025matpo,xu2026wideseek}.
Joint training also requires deciding how the team outcome
should guide each agent's updates.
In cooperative reinforcement learning, COMA uses counterfactual
baselines and QMIX factorizes team values
\citep{foerster2018coma,rashid2020qmix}.
Lin et al.\ use LLM-generated rewards to provide agent-specific
feedback \citep{lin2025teamwork}.
For collaborating language agents, CCPO studies counterfactual
and self-evaluated contributions \citep{li2026ccpo}.
Our focus is on conflicts between the team outcome and the utility
of individual subagent reports: a successful rollout may contain
misleading reports, and an unsuccessful one may contain useful
evidence. Rather than estimating a report's causal effect on the
final answer, LEAD-RL judges its evidential utility and uses that
judgment to gate the inherited team advantage. Helpful reports
do not inherit negative advantages, harmful reports do not inherit
positive advantages, and neutral reports retain the team signal.
Final-answer correctness remains the scalar team reward.

\begin{figure}[t]
    \centering
    \includegraphics[width=\textwidth]
    {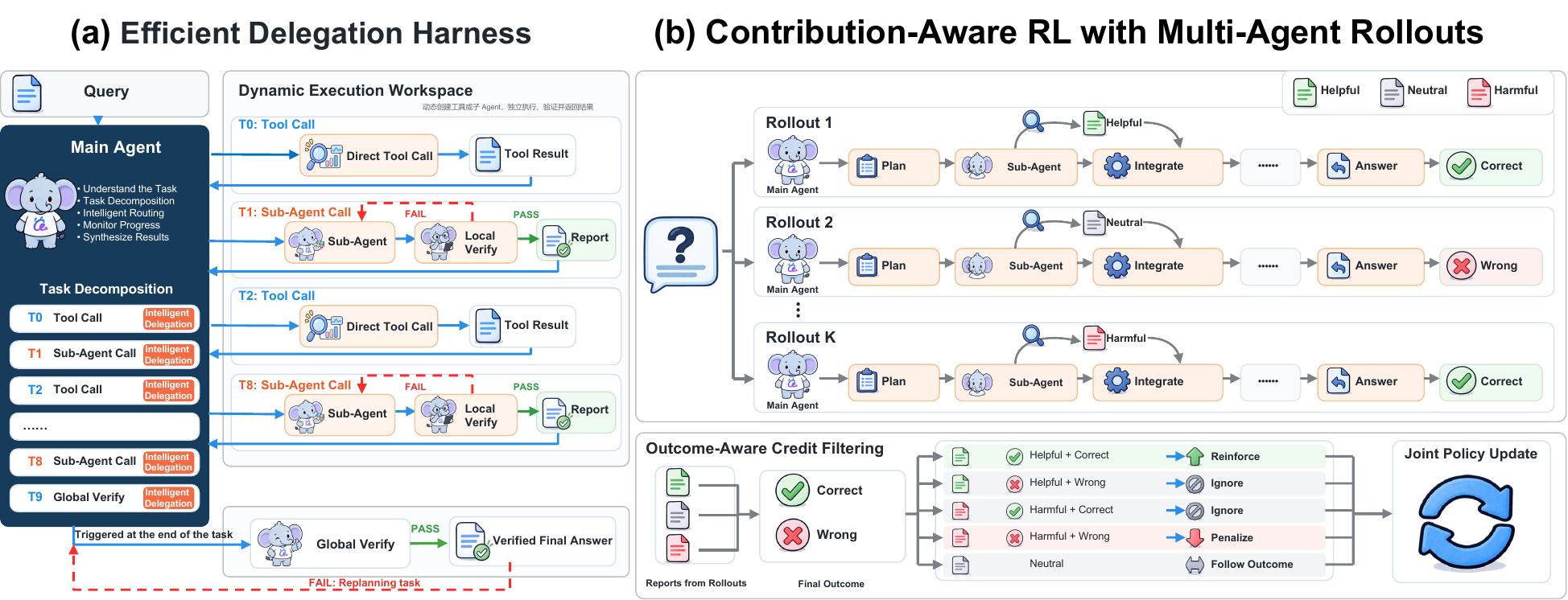}
    \caption{Overview of \textsc{LEAD-RL}. \textbf{(a)} The efficient
    delegation harness combines dynamic task routing with local and global
    verification. \textbf{(b)} Contribution-aware RL filters team-level
    outcome credit using \YangLiu{subagent report} judgments and jointly updates the
    \YangLiu{main agent} and \YangLiu{subagents}.}
    \label{fig:framework}
\end{figure}

\section{Method}
\label{sec:method}
\paragraph{Overview.}

Figure~\ref{fig:framework} summarizes \textsc{LEAD-RL}. The harness
supports direct tool use, parallel \YangLiu{subagents}, and evidence verification,
allowing the policy to distribute research across contexts. Training
jointly updates both roles from complete swarm rollouts, routing team
advantages according to \YangLiu{subagent report} contributions. Critical-path cost
and idealized speedup characterize the learned behavior but are not
reward terms; supervision remains anchored to final task outcomes.

\subsection{Problem Formulation and Efficiency Measures}
\label{sec:formulation}

\paragraph{Deep research with parallel delegation.}
We consider a long-horizon research task in which an agent answers a
question $q$ through multi-turn reasoning and tool use. Following the
main-distributes, sub-executes paradigm \citep{ning2026searchswarm}, the
system consists of a \YangLiu{main agent} and a dynamically instantiated set of
\YangLiu{subagents}. All roles are generated by a shared policy $\pi_\theta$ under
role-specific instructions.

The \YangLiu{main agent} maintains the global research state, uses tools directly,
and invokes $\texttt{call\_sub\_agent}(B_g)$ with briefs
$B_g=\{b_{g,1},\ldots,b_{g,n_g}\}$ at event $g$ to dispatch
$n_g$ \YangLiu{subagents} in parallel. \YangLiu{Subagent} $i$
executes a trajectory $H_{g,i}^S$ in an isolated context and returns a \YangLiu{subagent report}
$r_{g,i}$ to the \YangLiu{main agent}. Section~\ref{sec:harness} describes the brief,
\YangLiu{subagent report}, and verification protocol governing this interaction.

A complete rollout is $H=(H^M,\{H_{g,i}^S\}_{g,i},y)$, where $H^M$
is the trajectory of the \YangLiu{main agent} and $y$ its final answer, scored by the
task evaluator $R_{\mathrm{perf}}(q,y)$.

\paragraph{Critical steps.}
Let $\mathcal{G}(H)$ denote the set of delegation events in trajectory
$H$. For each event $g \in \mathcal{G}(H)$, let $n_g$ be the number of
\YangLiu{subagents} dispatched in parallel, and let $N_{g,i}^S(H)$ denote the
number of non-report tool-call steps taken by its $i$-th \YangLiu{subagent}. Let
$N_M(H)$ denote the total number of tool-call steps taken by the \YangLiu{main
agent}, where each invocation of \texttt{call\_sub\_agent} is counted as
one step taken by the \YangLiu{main agent}.

If all agent actions were executed sequentially, the trajectory would
require
\begin{equation}
    C_{\mathrm{seq}}(H)
    =
    N_M(H)
    +
    \sum_{g \in \mathcal{G}(H)}
    \sum_{i=1}^{n_g}
    N_{g,i}^S(H)
    \label{eq:sequential_steps}
\end{equation}
tool-call steps. We refer to this quantity as the
\emph{sequential-equivalent cost}, which measures the total amount of
work performed across the swarm.

\YangLiu{Subagents} belonging to the same delegation event execute concurrently.
The \YangLiu{main agent} resumes after their \YangLiu{subagent reports} return, so the effective cost of
an event is determined by its longest \YangLiu{subagent} branch. Following the
critical-path abstraction used for parallel agent systems
\citep{team2026kimi}, we define
\begin{equation}
    C_{\mathrm{crit}}(H)
    =
    N_M(H)
    +
    \sum_{g \in \mathcal{G}(H)}
    \max_{1 \leq i \leq n_g}
    N_{g,i}^S(H).
    \label{eq:critical_steps}
\end{equation}

Thus, delegation efficiency depends on the longest \YangLiu{subagent} branch:
launching more \YangLiu{subagents} does not shorten an unbalanced critical path.

\paragraph{Parallel speedup.}
We quantify the benefit obtained from parallel delegation using the
idealized speedup ratio
\begin{equation}
    S(H)
    =
    \frac{
        C_{\mathrm{seq}}(H)
    }{
        C_{\mathrm{crit}}(H)
    }.
    \label{eq:speedup}
\end{equation}
The ratio compares sequential and parallel execution of the same
workload, with $S(H)=1$ when no parallelism is used. It measures
step-level parallelism, not wall-clock acceleration: heterogeneous tool
latency, communication overhead, and resource contention are abstracted away.

\paragraph{Efficiency as an evaluation target.}
Critical-path cost supports cross-system comparison, while speedup
measures parallelization of a system's own workload. Since redundant
\YangLiu{subagents} can inflate speedup, we interpret it alongside answer quality,
critical-path cost, and total tokens. These measures evaluate whether
outcome-only learning improves the accuracy--efficiency trade-off;
they are not reward terms requiring hand-designed cost coefficients.

\subsection{\YangLiu{Efficient Delegation Harness}}
\label{sec:harness}

\YangLiu{Figure~\ref{fig:framework}(a) illustrates the interaction protocol used
to collect swarm rollouts and execute the learned policy. At each step, the
main agent chooses between direct tool use and parallel subagent calls. A local
verifier filters each report before it returns, and a global verifier checks the
synthesized answer before submission. The harness defines these interfaces;
the shared policy controls routing, search, and synthesis.}

\paragraph{Dynamic task routing.}
\YangLiu{The main agent routes each information need to direct search or
delegation based on the accumulated evidence, batching independent needs into
parallel calls. Each brief contains a bounded objective, relevant evidence, and
unresolved constraints, so the subagent does not require the full main-agent
history. The policy generates this routing rather than following fixed stages.}

\paragraph{Isolated execution and evidence return.}
\YangLiu{Each subagent searches in an isolated context and produces a report
with findings, citations, and remaining uncertainty. Its raw trajectory stays
isolated; only the verified report returns, whereas direct tool observations
enter the main context. Delegation thus controls which retrieval histories are
condensed into the global research state.}

\paragraph{\YangLiu{Local and global verification.}}
\YangLiu{A local verifier checks each candidate report for brief coverage and
evidential support. A passing report returns to the main agent; a failed check
sends feedback for additional research or revision.}
\YangLiu{At answer time, a global verifier checks the synthesized response for
consequential claims, citations, and task constraints. A passing response is
released; a failed check returns control for further research. Both verifiers
are role-conditioned invocations of the shared policy, use no reference answer
at inference, and are distinct from the training-time contribution rubric in
Figure~\ref{fig:framework}(b).}

\subsection{Contribution-Aware \YangLiu{Joint Main-Agent--Subagent Training}}
\label{sec:joint_rl}

Efficient delegation requires the \YangLiu{main agent}'s task allocation and
the \YangLiu{subagents}' search behavior to improve together. \YangLiu{Kimi K2.5,
for example, optimizes the main agent while holding subagents fixed}
\citep{team2026kimi}. Besides preventing co-adaptation
between the \YangLiu{main agent} and \YangLiu{subagents}, this can create a role-policy mismatch when a shared updated
checkpoint is used for both roles at inference. We instead optimize both
roles from the same swarm rollouts and route \YangLiu{subagent} credit according to the
contribution of each returned \YangLiu{subagent report}.

\paragraph{\YangLiu{Joint main-agent--subagent training}.}
The \YangLiu{main agent} and \YangLiu{subagents} share a policy $\pi_\theta$, conditioned
on their respective role contexts. For each question $q$, we sample
$K$ complete rollouts $H_j=(H_j^M,\{H_{j,i}^S\}_{i=1}^{n_j},y_j)$,
$j=1,\ldots,K$, where $H_j^M$ contains the \YangLiu{main agent}'s reasoning, delegation, and
synthesis, and $H_{j,i}^S$ records a \YangLiu{subagent}'s search and reporting.
Both roles are sampled from the same evolving policy and included in the
same policy update. The policy governing task allocation therefore adapts
alongside the policy responsible for \YangLiu{subagent} search and reporting.

The roles shape each other's training data: the \YangLiu{main agent} controls
subtask scope and requested evidence, while \YangLiu{subagents} determine the \YangLiu{subagent reports}
returned for synthesis. Updating both together avoids training the \YangLiu{main agent}
against permanently frozen \YangLiu{subagents} and then deploying an updated shared
checkpoint in both roles.

Role-specific instructions and contexts induce distinct conditional
behaviors despite parameter sharing. Each complete interaction of the
\YangLiu{main agent} or a \YangLiu{subagent} forms one logical training trajectory, regardless of its
number of model--tool turns. Tool observations provide context but receive
no language-model loss, preserving credit at the \YangLiu{subagent report} level.

Each rollout receives a team reward
$R_j=R_{\mathrm{perf}}(q,y_j)$ based only on its final answer. There is no
bonus for delegation, rubric labels, shorter trajectories, or higher
speedup. For the $K$ rollouts of the same question, we compute
the group-relative advantage
\begin{equation}
    A_j^{\mathrm{team}}
    =
    \frac{
        R_j-\operatorname{mean}_{k}R_k
    }{
        \operatorname{std}_{k}R_k+\epsilon
    }.
    \label{eq:team_advantage}
\end{equation}
The \YangLiu{main agent} receives $A_j^M=A_j^{\mathrm{team}}$, supervising
its delegation decisions through the overall research outcome.
\YangLiu{Subagents}, however, should not indiscriminately inherit the same
signal: a useful \YangLiu{subagent report} may appear in an unsuccessful rollout,
while a misleading \YangLiu{subagent report} may be corrected by a successful
\YangLiu{main agent}. We therefore assign \YangLiu{subagent} credit according to its
estimated contribution.

\paragraph{Dynamic \YangLiu{subagent} contribution monitoring.}
Separate from the final-answer judge that supplies the team reward,
a contribution judge assesses each \YangLiu{subagent report}. It examines the question $q$, reference
answer $y^\star$, delegated brief $b_{j,i}$, returned \YangLiu{subagent report}
$r_{j,i}$, and supporting search evidence $E_{j,i}$:
\begin{equation}
    c_{j,i}
    =
    \mathcal{J}
    \left(
        q,y^\star,b_{j,i},r_{j,i},E_{j,i}
    \right)
    \in\{+1,0,-1\}.
    \label{eq:contribution_assessment}
\end{equation}
A \emph{helpful} contribution ($+1$) provides accurate,
evidence-supported information that advances the research, such
as resolving a key fact, eliminating a plausible alternative, or
clarifying an ambiguity. A \emph{harmful} contribution ($-1$)
introduces materially misleading conclusions or unsupported
claims. Inconclusive or uncertain contributions are assigned
\emph{neutral} credit ($0$).

The evidence consists only of tool outputs captured from that \YangLiu{subagent}'s
execution, not additional searches performed by the judge. Neither the
\YangLiu{main agent}'s final response nor its correctness verdict is included in
the contribution-judging input. The assessment therefore concerns the
\YangLiu{subagent report}'s potential research utility, rather than whether the team
ultimately succeeded.

The rubric assesses the relevance and evidential quality of each
\YangLiu{subagent}'s output, rather than merely whether it completed its assignment or
whether the final team answer happened to be correct. A \YangLiu{subagent report} may therefore
be helpful without stating the reference answer---for example, by
evidence-groundedly eliminating a plausible candidate---and an apparently
relevant \YangLiu{subagent report} remains neutral when its support or downstream utility cannot
be established. The judge returns a label, a confidence score, the decisive
claim, and a short justification. We use a conservative decision policy:
helpful or harmful judgments with confidence below $0.8$ are mapped to
neutral, while materially misleading
claims are required for a harmful label. This reduces the risk that rubric
noise suppresses valid \YangLiu{subagent} behavior.

Assessment is repeated for newly sampled \YangLiu{subagent reports} throughout training, rather
than being assigned once to a fixed data set. The contribution signal thus
tracks changes in both the briefs produced by the \YangLiu{main agent} and the \YangLiu{subagent reports}
produced by \YangLiu{subagents}. The judge is used only for credit assignment during
training; neither it nor the reference answer is available to the agents at
inference time. The complete rubric is provided in
Appendix~\ref{app:rubric-training}.

\paragraph{Contribution-aware credit assignment.}
We route the team advantage through a direction-consistent gate:
\begin{equation}
    g(c,A)
    =
    \begin{cases}
        \max(A,0), & c=+1,\\
        A,         & c=0,\\
        \min(A,0), & c=-1.
    \end{cases}
    \label{eq:credit_gate}
\end{equation}
Helpful \YangLiu{subagents} are protected from negative team advantages, while
harmful \YangLiu{subagents} cannot receive positive team advantages. Neutral \YangLiu{subagents}
retain outcome supervision, preserving learning where individual utility
is uncertain. The rubric therefore intervenes only when the direction of
\YangLiu{subagent} credit is sufficiently clear.

The resulting \YangLiu{subagent} advantage is
\begin{equation}
    A_{j,i}^S
    =
    \frac{w(t)}
         {\max(1,|\mathcal{I}_j|)}
    g(c_{j,i},A_j^{\mathrm{team}}),
    \label{eq:worker_advantage}
\end{equation}
where $\mathcal{I}_j$ contains trainable \YangLiu{subagents} with nonzero gated
advantages, and $w(t)$ controls the relative weight of \YangLiu{subagent} learning and
is annealed during training. Normalization prevents rollouts with more \YangLiu{subagents}
from receiving disproportionately large aggregate updates.

Let $\ell_{\mathrm{GSPO}}(h,A)$ denote the clipped policy loss of
Group Sequence Policy Optimization (GSPO) \citep{zheng2025gspo}
for trajectory $h$ with advantage $A$. We jointly optimize
\begin{equation}
    \mathcal{L}_{\mathrm{joint}}
    =
    \mathbb{E}
    \left[
        \frac{1}{K}\sum_{j=1}^{K}
        \left(
            \ell_{\mathrm{GSPO}}
            (H_j^M,A_j^{\mathrm{team}})
            +
            \sum_{i=1}^{n_j}
            \ell_{\mathrm{GSPO}}
            (H_{j,i}^S,A_{j,i}^S)
        \right)
    \right]
    +
    \beta\mathcal{L}_{\mathrm{KL}}.
    \label{eq:joint_rl}
\end{equation}

The KL term anchors both roles to the same reference policy.
Normalized \YangLiu{subagent} advantages prevent additional branches from
automatically increasing the aggregate \YangLiu{subagent} update. The objective
rewards neither delegation frequency nor a prescribed swarm topology.

Appendix~\ref{app:rubric-training} summarizes the rollout, judging,
credit-assignment, and joint-update procedure.

\paragraph{Why outcome-only learning can improve efficiency.}
LEAD-RL corrects two supervision errors: penalizing helpful \YangLiu{subagents} for
team failures and reinforcing harmful \YangLiu{subagents} for team successes.
More reliable \YangLiu{subagent reports} may reduce compensatory searches and support more
selective delegation, while normalized \YangLiu{subagent} credit avoids scaling
updates with \YangLiu{subagent} count. This is a hypothesis about learned behavior,
not a guarantee of minimum cost. We examine it through
accuracy--efficiency trajectories and comparisons with main-only and
shared-team-credit RL, without prescribing a trade-off between
correctness and execution cost.

\section{Experiments}
\label{sec:experiments}

\subsection{Experimental Setup}
\label{sec:experimental_setup}

\paragraph{Benchmarks.}
We evaluate LEAD-RL on BrowseComp \citep{wei2025browsecomp} and
BrowseComp-ZH \citep{zhou2025browsecomp} for English and Chinese factual
search, GAIA \citep{mialon2024gaia} for general assistant tasks, Humanity's
Last Exam (HLE) \mbox{\citep{hle2026}} for expert questions, and
DeepSearchQA \citep{gupta2026deepsearchqa} for set-valued research. We use
the full BrowseComp benchmark and tool-augmented HLE. DeepSeek V4 Pro judges
our runs and validation checkpoints; external scores retain their source
evaluation protocols.

\paragraph{Evaluation metrics.}
We report final-answer accuracy for BrowseComp, BrowseComp-ZH, GAIA, and
HLE, and mean per-question answer-set $F_1$ for DeepSearchQA; all are
percentages. Efficiency metrics are total \YangLiu{main agent} and \YangLiu{subagent} tokens per
question (\emph{Tokens/Q}), mean critical-path cost ($C_{\mathrm{crit}}/Q$;
Equation~\ref{eq:critical_steps}), and idealized speedup
(Equation~\ref{eq:speedup}). We also report rubric-assessed helpful and
harmful \YangLiu{subagent report} rates.

\paragraph{Baselines.}
Table~\ref{tab:main-results} compares closed-source systems, large
open-source models, and lightweight research agents; Appendix~\ref{app:baseline-details}
lists their references.
External scores come from technical reports or model cards and may use
different scaffolds, so they provide context rather than a controlled
training comparison.

Within the same swarm harness, we evaluate Qwen3.8-Max
\citep{qwen2026qwen38}, DeepSeek V4 Pro and
DeepSeek V4 Flash \citep{deepseek2026v4}, Ling-3-Flash
\citep{inclusionai2026ling3flash}, and Qwen3.6-35B-A3B
\citep{qwen2026qwen36}. Matched base-versus-trained and swarm-on/off runs
isolate training and delegation effects.

\paragraph{Training variants.}
We compare four training variants:
\emph{SFT initialization};
\emph{Main-only RL}, which keeps the subagents fixed;
\emph{joint RL with shared team credit}, which broadcasts the
team advantage to both roles; and
\emph{LEAD-RL}, which gates subagent advantages using
rubric-based report labels.
All three RL variants start from the same SFT checkpoint and
are trained on the same data. We select checkpoints using
the same validation set.

\paragraph{Implementation details.}
We use sample-level GSPO with $K=4$ rollouts per question.
DeepSeek V4 Pro provides final-answer and contribution judgments using
separate temperature-$0$ prompts whose outputs never enter agent contexts.
Correctness is the sole team reward; rubric labels only route \YangLiu{subagent}
credit. Appendix~\ref{app:rubric-training} gives the judging protocol and
hyperparameters. Controlled comparisons share the question split, decoding
configuration, harness, and execution budget.

\subsection{Performance across Benchmarks and Model Families}
\label{sec:main_results}

\begin{table}[t]
\centering
\caption{\YangLiu{Main results across five deep-research benchmarks. DeepSearchQA
reports mean answer-set $F_1$; all other columns report accuracy.
$^{*}$ denotes context management, $^{\dagger}$ denotes runs executed in our
swarm harness, and blue rows denote LEAD-RL.}}

\label{tab:main-results}
\resizebox{\linewidth}{!}{%
\begin{tabular}{lcccccc}
\toprule
\textbf{Model} & \textbf{Size} & \textbf{BrowseComp} & \textbf{BrowseComp-ZH} & \textbf{GAIA} & \textbf{HLE (w/ Tools)} & \textbf{DeepSearchQA ($F_1$)} \\
\midrule
\rowcolor{gray!8}
\multicolumn{7}{l}{\textit{Closed-source models}} \\
GPT-5 & -- & 54.9 & 65.0 & 76.4 & 35.2 & 79.0 \\
Claude-4.5-Sonnet & -- & 24.1 & 42.4 & 66.0 & 32.0 & 27.9 \\

\midrule
\rowcolor{gray!8}
\multicolumn{7}{l}{\textit{Open-source models}} \\
GLM-4.7 & 355B-A32B & 67.5\textsuperscript{*} & 66.6\textsuperscript{*} & -- & 42.8 & -- \\
DeepSeek V3.2 & 671B-A37B & 67.6\textsuperscript{*} & 65.0\textsuperscript{*} & 75.1 & 40.8 & 60.9 \\
LongCat-Flash-Thinking-2601 & 560B-A27B & 73.1\textsuperscript{*} & 77.7\textsuperscript{*} & -- & -- & -- \\
MiniMax-M2 & 230B-A10B & 44.0 & -- & 75.7 & -- & -- \\
Step-3.5-Flash & 196B-A11B & 69.0\textsuperscript{*} & 66.9 & 84.5 & -- & -- \\

\midrule
\rowcolor{gray!8}
\multicolumn{7}{l}{\textit{Open-source lightweight models}} \\
RedSearcher & 30B-A3B & 57.4\textsuperscript{*} & 58.2\textsuperscript{*} & 80.1 & 34.3 & -- \\
LongSeeker & 30B-A3B & 61.5\textsuperscript{*} & 62.5\textsuperscript{*} & 77.7\textsuperscript{*} & -- & -- \\
MiroThinker-1.5-mini & 30B-A3B & 56.1\textsuperscript{*} & 66.8\textsuperscript{*} & 72.0\textsuperscript{*} & 31.0\textsuperscript{*} & -- \\
MiroThinker-1.7-mini & 30B-A3B & 67.9\textsuperscript{*} & 72.3\textsuperscript{*} & 80.3\textsuperscript{*} & 36.4\textsuperscript{*} & 67.9\textsuperscript{*} \\

\midrule
\rowcolor{gray!8}
\multicolumn{7}{l}{\textit{\YangLiu{Models evaluated in our swarm harness}}} \\
\YangLiu{Qwen3.8-Max$^{\dagger}$} & 2.4T-A95B & 72.3 & 73.4 & 92.1 & -- & -- \\
\YangLiu{DeepSeek V4 Pro$^{\dagger}$} & \YangLiu{1.6T-A49B} & 75.3 & 74.7 & 97.0 & 62.2 & 93.5 \\
\YangLiu{DeepSeek V4 Flash$^{\dagger}$} & \YangLiu{284B-A13B} & 71.2 & 73.4 & 92.2 & 58.8 & 91.9 \\
\YangLiu{Ling-3-Flash$^{\dagger}$} & 124B-A5.1B & 58.0 & 71.3 & 73.1 & 30.8 & 80.2 \\
\YangLiu{Qwen3.6-35B-A3B$^{\dagger}$} & 35B-A3B & 58.3 & 70.9 & 81.6 & 41.7 & 83.2 \\

\midrule
\multicolumn{7}{l}{\textit{\YangLiu{Baseline: Ling-3-Flash}}} \\
\rowcolor{blue!5}
\textbf{\YangLiu{LEAD-RL (Ling-3-Flash)$^{\dagger}$}} & 124B-A5.1B & \textbf{73.5} & \textbf{76.3} & \textbf{94.2} & \textbf{56.2} & \textbf{86.3} \\

\multicolumn{7}{l}{\textit{\YangLiu{Baseline: Qwen3.6-35B-A3B}}} \\
\rowcolor{blue!5}
\textbf{\YangLiu{LEAD-RL (Qwen3.6-35B-A3B)$^{\dagger}$}} & 35B-A3B & \textbf{70.0} & \textbf{72.9} & \textbf{88.4} & \textbf{54.3} & \textbf{85.2} \\
\bottomrule
\end{tabular}%
}
\end{table}

Table~\ref{tab:main-results} shows gains across both model families and
all five benchmarks, spanning factual search, general tool use, and
set-valued research. This breadth suggests improvements beyond a single
answer format or architecture. These comparisons assess the complete pipeline;
Section~\ref{sec:credit_ablation} isolates RL relative to SFT.

The Qwen-based model also surpasses the reported MiroThinker-1.7-mini
results across all five benchmarks without additional context-management
strategies beyond swarm delegation. This comparison with a similarly
sized, context-managed model supports distributing research across \YangLiu{subagent}
contexts as an effective approach to long-horizon tasks. External
evaluation protocols differ; the matched comparisons below examine
learned delegation more directly.

\subsection{Accuracy--Efficiency Trade-offs under Swarm Execution}
\label{sec:swarm_effect}

\begin{table}[t]
\centering
\caption{Effect of swarm execution on BrowseComp.
Accuracy columns are matched runs under the same budget; efficiency is
reported for swarm-enabled runs.}
\label{tab:harness-browsecomp}
\resizebox{\linewidth}{!}{%
\begin{tabular}{lcccccc}
\toprule
& & \multicolumn{2}{c}{\textbf{Accuracy (\%) $\uparrow$}} &
\multicolumn{3}{c}{\textbf{Swarm efficiency}} \\
\cmidrule(lr){3-4} \cmidrule(lr){5-7}
\textbf{Model} & \textbf{Size} & \textbf{w/o Swarm} &
\textbf{w/ Swarm} & \textbf{Tokens/Q (k) $\downarrow$} &
\textbf{$C_{\mathrm{crit}}$/Q $\downarrow$} & \textbf{Speedup $\uparrow$} \\
\midrule
Qwen3.8-Max & 2.4T-A95B & 63.2 & 72.3 & 277.3 & 159.09 & 1.67x \\
DeepSeek V4 Pro & 1.6T-A49B & 67.4 & 75.3 & 350.1 & 266.13 & 1.39x \\
DeepSeek V4 Flash & 284B-A13B & 65.2 & 71.2 & 401.7 & 293.80 & 1.44x  \\
Ling-3-Flash & 124B-A5.1B & 59.0 & 58.0 & 330.7 & 251.0 & 1.69x \\
\midrule
\rowcolor{blue!5}
\textbf{LEAD-RL (Ours)}& 124B-A5.1B  & 66.5 & 73.5 & 336.9 & 244.22 & 1.71x \\
\bottomrule
\end{tabular}%
}
\label{tab:swarm-effect}
\end{table}

Table~\ref{tab:swarm-effect} compares the effect of swarm execution
before and after training. Enabling the swarm changes the base Ling
model's accuracy from 59.0\% to 58.0\%, whereas LEAD-RL improves from
66.5\% to 73.5\%. With the swarm enabled, training raises accuracy
from 58.0\% to 73.5\%. Token consumption increases slightly from
330.7k to 336.9k tokens per question, while critical-path cost
decreases from 251.0 to 244.22 steps per question. The trained model
therefore gains more from delegation without a substantial increase
in token consumption.

This improvement makes a substantially smaller model competitive with
larger systems in the same harness. LEAD-RL (124B-A5.1B) outperforms
DeepSeek V4 Flash (284B-A13B) while using fewer tokens and critical
steps, despite having less than half its total and active parameters.
It also reaches 73.5\% accuracy, within 1.8 percentage points of the
1.6T-A49B DeepSeek V4 Pro, with lower measured execution costs. These
comparisons show that training effective delegation can narrow the
performance gap to much larger models while retaining a compact
parameter footprint.

Qwen3.8-Max uses fewer tokens and has a shorter critical path than
LEAD-RL. LEAD-RL achieves 73.5\% accuracy on BrowseComp, 1.2 percentage
points higher than Qwen3.8-Max, with substantially fewer total and
active parameters.

\subsection{Effect of Contribution-Aware \YangLiu{Joint Main-Agent--Subagent Training}}
\label{sec:credit_ablation}






\begin{table}[t]
\centering
\caption{Training variants on BrowseComp.
Main-only RL freezes subagents, while joint RL with shared credit
updates both roles using the team advantage. LEAD-RL gates subagent
advantages using report utility labels.
Helpful and Harmful denote judge-assessed report rates.
$C_{\mathrm{crit}}/Q$ measures critical-path steps per question;
Speedup measures idealized step-level parallelism.}
\label{tab:rl-ablation}

\small
\setlength{\tabcolsep}{4pt}
\renewcommand{\arraystretch}{1.15}

\begin{tabular*}{\linewidth}{
    @{\extracolsep{\fill}}lccccc@{}
}
\toprule
&
& \multicolumn{2}{c}{\textbf{Report labels (\%)}}
& \multicolumn{2}{c}{\textbf{Efficiency}} \\
\cmidrule(lr){3-4}
\cmidrule(l){5-6}
\textbf{Training variant}
& \textbf{Acc. (\%)} $\uparrow$
& \textbf{Helpful} $\uparrow$
& \textbf{Harmful} $\downarrow$
& $C_{\mathrm{crit}}/Q$ $\downarrow$
& \textbf{Speedup} \\
\midrule

SFT initialization
& 72.50 & 27.11 & 12.89 & 403.37 & $1.35\times$ \\

\midrule

Main-only RL
& 70.39 & 27.42 & 11.21 & 421.76 & $1.42\times$ \\

Joint RL (shared credit)
& 71.00 & 21.15 & 13.84 & 323.30 & $1.39\times$ \\

\rowcolor{blue!5}
\textbf{LEAD-RL (ours)}
& \textbf{73.49}
& \textbf{31.52}
& \textbf{9.87}
& \textbf{244.22}
& $\mathbf{1.71}\times$ \\

\bottomrule
\end{tabular*}
\end{table}

Table~\ref{tab:rl-ablation} compares four training variants on
BrowseComp. Main-only RL and joint RL with shared team credit both
yield lower accuracy than the SFT checkpoint. LEAD-RL achieves
73.49\% accuracy, exceeding joint RL with shared team credit by
2.49 percentage points. It also reduces critical-path cost from
323.30 to 244.22 steps per question, a decrease of 24.5\%.

LEAD-RL also produces more rubric-assessed helpful \YangLiu{subagent reports} and fewer
harmful ones than shared team credit. Relative to SFT, it improves accuracy
while reducing critical-path cost by 39.5\%. More reliable reports may reduce
corrective search, although we do not measure that mechanism directly. The
gains therefore extend beyond recovering performance lost under other RL
variants and arise without an explicit cost penalty.

\subsection{Learning Dynamics without Explicit Cost Rewards}
\label{sec:training_dynamics}

\begin{figure}[t]
    \centering
    \includegraphics[width=\textwidth]
    {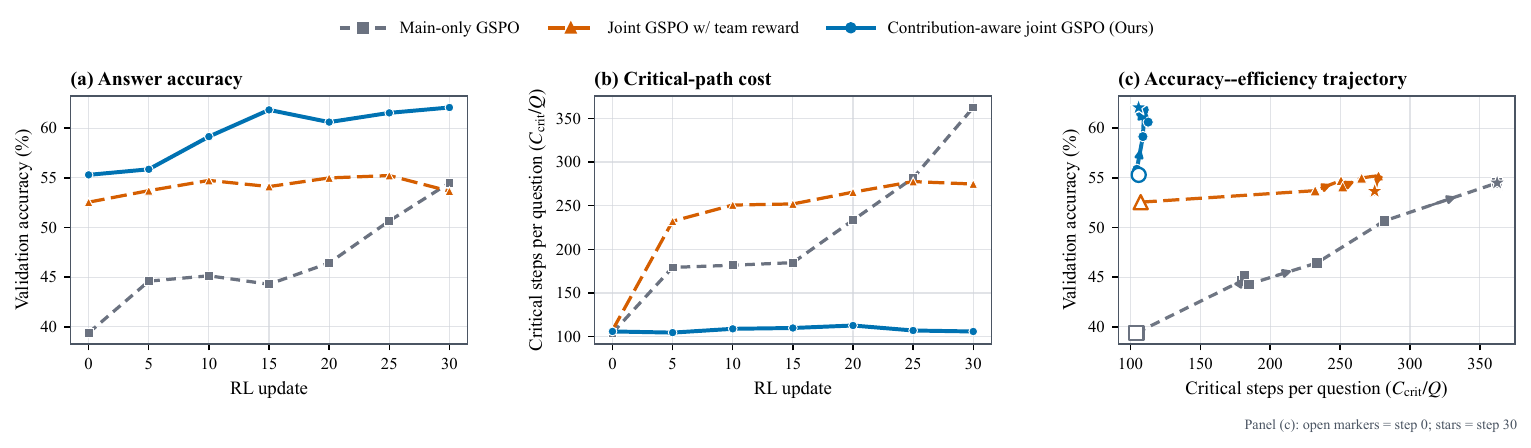}
    \caption{Validation dynamics across RL checkpoints on a fixed validation
    set. Open markers and stars denote the first and last recorded checkpoints;
    arrows indicate training direction.}
    \label{fig:training_dynamics}
\end{figure}

Figure~\ref{fig:training_dynamics} compares checkpoints on a fixed
validation set; step~0 follows the first RL update, so comparisons are
within-run over the common recorded horizon.

Main-only and shared-team-credit training develop substantially longer
critical paths; the latter gains little accuracy despite this growth.
LEAD-RL instead gains 6.78 accuracy points at approximately constant
critical-path cost, producing the upward trajectory in panel~(c), even
though cost is absent from the reward. The favorable trade-off thus develops
during training rather than appearing only at the final checkpoint.
Appendix~\ref{app:raw-training-dynamics}
reports the raw on-policy measurements; later checkpoints remain stable
within the observed horizon.

Rubric-guided filtering offers one explanation: blocking positive updates
to harmful \YangLiu{subagent reports} and negative updates to helpful ones reduces conflicting
supervision. This may promote more useful delegated research and fewer
unnecessary follow-up searches, but the mechanism remains a hypothesis
rather than a directly measured effect.

\section{Conclusion}
\label{sec:conclusion}

We presented \textsc{LEAD-RL}, which learns delegation as active context
management through \YangLiu{joint main-agent--subagent training} and rubric-based
credit assignment. Final-answer correctness remains the sole team reward.
Across five benchmarks and two model families, the trained models improve
answer quality and compete with substantially larger systems. Controlled
RL comparisons and validation dynamics show improved accuracy--efficiency
trade-offs without explicit cost penalties. These findings suggest that
correcting how \YangLiu{subagents} inherit outcome credit can improve both research
quality and the use of computation in agent swarms.


\bibliographystyle{assets/plainnat}
\bibliography{main}

\clearpage
\appendix

\section{Contribution Rubric and \YangLiu{Joint Main-Agent--Subagent Training}}
\label{app:rubric-training}

\paragraph{Two distinct judging tasks.}
We use DeepSeek V4 Pro as an external judge for two training signals,
with temperature $0$ and a maximum completion budget of $16{,}384$ tokens
per request. The final-answer judge receives the original question,
reference answer, and \YangLiu{main agent} response. Following the BrowseComp
judging protocol, it extracts the final answer and checks semantic
equivalence to the reference, allowing a small numerical tolerance where
appropriate. Its parsed \texttt{correct: yes/no} verdict determines the
binary team reward. The contribution judge instead evaluates individual
\YangLiu{subagent reports}; its labels only route the resulting team advantage and
never alter that reward.

\paragraph{Contribution-judge inputs.}
Each item contains the original question, reference answer, delegated
brief, \YangLiu{subagent report}, and evidence captured from that \YangLiu{subagent}'s tool
calls. No \YangLiu{main agent} final response or team-correctness label is included.
Evidence entries retain the tool name, arguments, and returned content.
We prioritize visited-page content over search results, preserving the
original order within each category. The judge does not browse for
additional evidence or infer unobserved tool calls. The question,
reference answer, brief, \YangLiu{subagent report}, and evidence use character budgets of
$4{,}000$, $2{,}000$, $4{,}000$, $6{,}000$, and $8{,}000$, respectively;
truncated fields are explicitly marked. Evidence may be empty, in which
case useful claims still require support in the supplied inputs.

\paragraph{Contribution rubric.}
The rubric evaluates potential research utility rather than downstream
use by the \YangLiu{main agent}. Supported elimination of a plausible candidate
can be helpful even without stating the final answer; an unsupported
answer guess is neutral, even if it matches the reference. Harmful
labels require a materially misleading assertion, not merely a failed
or incomplete search.
The complete judge instruction is reproduced below.

\begin{tcolorbox}[
  enhanced,
  breakable,
  colback=blue!2,
  colframe=blue!35!black,
  colbacktitle=blue!8,
  coltitle=black,
  title={Contribution-rubric judge prompt},
  fonttitle=\bfseries,
  boxrule=0.55pt,
  arc=1mm,
  left=1.2mm,
  right=1.2mm,
  top=1mm,
  bottom=1mm,
  before upper={\small\setlength{\parindent}{0pt}}
]
You are a strict but conservative grader of one SearchSwarm \YangLiu{subagent}'s
research contribution to solving a question.

\medskip
\textbf{For every item you receive:}
\begin{itemize}[leftmargin=1.6em,itemsep=0pt,topsep=2pt,parsep=0pt]
    \item \texttt{original\_question}
    \item \texttt{reference\_answer}
    \item \texttt{delegated\_task}
    \item \texttt{report}
    \item \texttt{evidence}: excerpts actually returned by the \YangLiu{subagent}'s
    tools; it may be empty
\end{itemize}

\textbf{Classify exactly one label:}
\begin{itemize}[leftmargin=1.6em,itemsep=2pt,topsep=2pt,parsep=0pt]
    \item \textbf{helpful:} The \YangLiu{subagent report} supplies accurate, non-trivial
    information that materially helps derive or verify the reference
    answer, OR it correctly rules out a plausible competing hypothesis or
    resolves an ambiguity in a way that materially narrows the search. The
    useful claim must be supported by the supplied \YangLiu{subagent report}/evidence, not
    merely guessed.

    \item \textbf{harmful:} The \YangLiu{subagent report} makes a confident, materially
    misleading claim that pushes the investigation toward a wrong answer
    or away from the reference answer. This includes confidently endorsing
    a wrong candidate, presenting weak extrapolation as established fact,
    contradicting decisive supplied evidence, or fabricating evidence.

    \item \textbf{neutral:} Everything else, including relevant but
    generic work, duplicated work, an unresolved candidate list, an
    unsupported guess, a cautious failed search, weakly supported
    speculation, or information whose contribution cannot be established
    from the supplied inputs.
\end{itemize}

\textbf{Rules:}
\begin{enumerate}[leftmargin=1.8em,itemsep=1pt,topsep=2pt,parsep=0pt]
    \item Judge potential research utility, not whether the \YangLiu{main agent} used
    the \YangLiu{subagent report} and not whether the \YangLiu{main agent}'s final response was correct.
    \item The \YangLiu{subagent report} need not literally contain the reference answer. A
    well-supported elimination of a plausible wrong hypothesis can be
    helpful.
    \item Do not infer missing searches, visits, citations, or facts.
    Prefer neutral when support or effect is unclear.
    \item Use harmful only for a material misleading assertion, not merely
    for an incomplete or unsuccessful investigation.
    \item The reference answer anchors the final target, but intermediate
    claims may still be useful if they form a valid path toward it.
    \item Be robust to spelling variants, translations, and aliases; judge
    semantics, not string equality.
    \item A \YangLiu{subagent report} that merely states the reference answer without
    supporting evidence is neutral.
\end{enumerate}

\textbf{Return JSON only:}

\begin{tcolorbox}[
  colback=white,
  colframe=blue!20!black,
  boxrule=0.3pt,
  arc=0.5mm,
  left=1mm,
  right=1mm,
  top=0.7mm,
  bottom=0.7mm
]
{\footnotesize\ttfamily
\{\char34 items\char34:[\{\char34 id\char34:\char34...\char34,
\char34 label\char34:\char34 helpful|neutral|harmful\char34,\par
\char34 confidence\char34:0.0,
\char34 decisive\_claim\char34:\char34...\char34,
\char34 reason\char34:\char34...\char34\}]\}
\par}
\end{tcolorbox}

Return exactly one result for every input id and no additional fields.
\end{tcolorbox}

\paragraph{Structured output and conservative filtering.}
Requests batch up to two \YangLiu{subagent reports}, each identified by an item ID. The
judge returns a JSON record containing the label, confidence, decisive
claim, and justification for every input ID. We validate the ID mapping
and label/confidence ranges. We treat the judge-reported confidence as a
filtering signal rather than a calibrated probability: helpful or harmful predictions below $0.8$ are
mapped to neutral. Confidence and textual justifications are not added
to the reward or used as continuous advantage weights; only the
effective three-way label controls credit assignment.

\paragraph{Incomplete trajectories and judge failures.}
\YangLiu{Subagents} that terminate because of model-side limits or fail to produce
a \YangLiu{subagent report} receive neutral credit without an LLM contribution judgment.
\YangLiu{Subagent} trajectories affected by infrastructure failures are excluded
from contribution-based \YangLiu{subagent} updates. This differs from temporary
unavailability of the auxiliary contribution judge: after bounded
retries, completed \YangLiu{subagents} fall back to neutral while the already
computed \YangLiu{main agent} reward is preserved. Neutral \YangLiu{subagents} with trainable
tokens retain ordinary team-advantage supervision; neutral does not
mean that their trajectories are discarded. Contribution judgments and
reference answers are used for training-side supervision, not returned
to the agents, and the contribution judge adds no inference-time
computation.

\paragraph{\YangLiu{Joint main-agent--subagent training} procedure.}
For each question, we sample $K=4$ complete swarm rollouts from the
current shared policy. Final-answer rewards are normalized within the
group to produce $A_j^{\mathrm{team}}$. The trajectory of the \YangLiu{main agent}
inherits this advantage directly. For each \YangLiu{subagent}, the rubric label is
used by the gate in Equation~\ref{eq:credit_gate}: helpful \YangLiu{subagent reports} retain
only positive team advantages, harmful \YangLiu{subagent reports} retain only negative
advantages, and neutral \YangLiu{subagent reports} retain the original advantage. The \YangLiu{main agent}
and retained \YangLiu{subagent} trajectories are then optimized together using the
joint objective in Equation~\ref{eq:joint_rl}. Thus, the rubric neither
changes the task reward nor rewards additional delegation; it only routes
the existing outcome signal across \YangLiu{subagent} trajectories.

\paragraph{Training hyperparameters.}
Each update contains 32 questions with $K=4$ complete rollouts per
question, yielding approximately 128 \YangLiu{main agent} trajectories and their
associated \YangLiu{subagent} trajectories. Both roles share model parameters and
use role-specific contexts. Each \YangLiu{main agent} rollout may instantiate up
to 16 \YangLiu{subagents}. We use sample-level GSPO with a KL penalty coefficient
of $0.001$ against the reference policy. The relative \YangLiu{subagent} weight
$w(t)$ is held at $0.4$ for the first $10\%$ of training, annealed to
$0.2$ by $70\%$, and held constant thereafter.

\section{Additional Baseline Details}
\label{app:baseline-details}

The closed-source baselines in Table~\ref{tab:main-results} include GPT-5
\citep{openai2025gpt5} and Claude-4.5-Sonnet
\citep{anthropic2025sonnet45}. Large open-source baselines include
GLM-4.7 \citep{glm2025glm47}, DeepSeek V3.2
\citep{liu2025deepseek}, LongCat-Flash-Thinking-2601
\citep{team2026longcat}, MiniMax-M2 \citep{minimax2025m2}, and
Step-3.5-Flash \citep{huang2026step}. At the 30B-A3B scale, we compare
against RedSearcher \citep{chu2026redsearcher}, LongSeeker
\citep{lu2026longseeker}, and MiroThinker-1.5/1.7-mini
\citep{team2026mirothinker}.

\section{Raw On-Policy Training Dynamics}
\label{app:raw-training-dynamics}

\begin{figure*}[t]
    \centering
    \includegraphics[width=\textwidth]
    {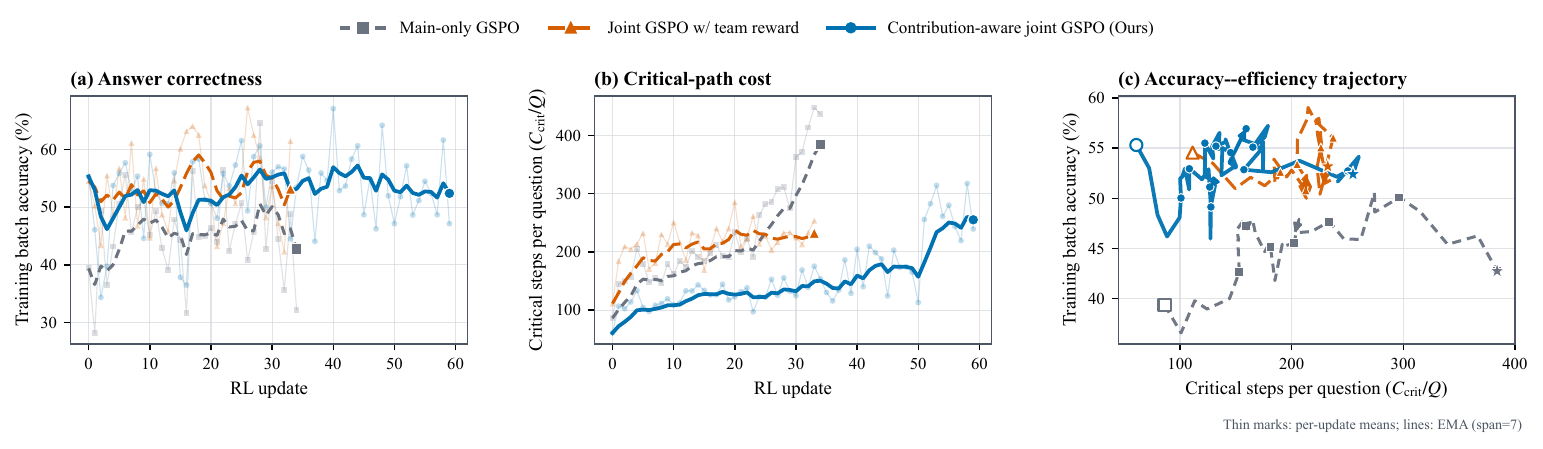}
    \caption{Raw on-policy training dynamics. Thin marks show per-update
    batch means and dark curves span-7 EMAs; open markers and stars denote
    the initial and final updates.}
    \label{fig:raw-training-dynamics}
\end{figure*}

Figure~\ref{fig:raw-training-dynamics} complements the fixed-validation
analysis in Section~\ref{sec:training_dynamics} with measurements from
the on-policy training batches. The main-only, shared-team-credit, and
LEAD-RL runs contain 35, 34, and 60 recorded updates, respectively. Each
thin mark is the mean of one sampled update batch; the dark curves use an
exponential moving average (EMA) with span seven. Because both questions
and trajectories vary across updates, these measurements expose the
behavior optimized during training but are not estimates of generalization
performance.

\paragraph{Outcome signal.}
Panel (a) is intentionally noisy because every update evaluates a newly
sampled set of questions and trajectories. Its EMA remains largely in the
$50$--$55\%$ range for the two joint-training variants, while main-only
GSPO rises from a lower initialization. The absence of a monotonic training
curve is therefore not evidence of optimization failure: group-relative
GSPO uses within-question reward differences, whereas the plotted batch
mean also reflects variation in batch difficulty. Fixed-set validation in
Figure~\ref{fig:training_dynamics} provides the controlled measure of
performance improvement.

\paragraph{Growth of the critical path.}
Panel (b) reveals a much clearer difference between the objectives. At
update 30, the EMA critical-path costs of main-only GSPO, joint GSPO with
shared team reward, and LEAD-RL are $296.45$, $226.18$, and $132.23$ steps
per question, respectively. LEAD-RL is therefore $55.4\%$ below main-only
training and $41.5\%$ below shared-team-credit training at the same update,
while their EMA batch correctness values are $50.09\%$, $55.06\%$, and
$55.19\%$. Thus, the baselines obtain their training progress by rapidly
expanding the effective research path, whereas contribution-aware joint
training remains substantially less search-intensive.

\paragraph{Accuracy--efficiency trajectory.}
Panel (c) combines the two signals. During the shared training horizon,
LEAD-RL occupies the upper-left region more consistently: it preserves
batch correctness while avoiding the pronounced rightward drift of the
baselines. In the longer LEAD-RL run, critical-path cost begins to rise
more rapidly after approximately 50 updates without a sustained increase
in batch correctness. This late-stage behavior illustrates why monitoring
efficiency throughout RL is necessary and why checkpoint selection should
consider the joint accuracy--efficiency frontier rather than defaulting to
the final optimization step.

\section{Case Study: Delegation as Active Context Management}
\label{app:case-study}

We examine one successful \textsc{LEAD-RL} trajectory to illustrate the
learned structure of delegation. The example requires matching several
independent clues against historical cricket scorecards. The \YangLiu{main agent}
first identifies five plausible matches, assigns one candidate to each of
five \YangLiu{subagents} in a single parallel dispatch, and then verifies and
synthesizes the returned evidence. For readability, we omit raw search
results and \YangLiu{subagent} scratchpads and show only the information that crosses
the delegation boundary. This qualitative example is not used in the
aggregate evaluation.

\noindent\textbf{User question.}\par\smallskip
\begin{tcolorbox}[
  colback=black!2,
  colframe=black!35,
  boxrule=0.45pt,
  arc=1mm,
  left=1.5mm,
  right=1.5mm,
  top=1mm,
  bottom=1mm,
  before upper={\small}
]
In a cricket match played between 1980 and 1990 (exclusive), a player was
dismissed in one of the rarest ways, with no credit given to the bowler. In
the same match, three players were dismissed for a duck, and three players
were run out. The team that won the toss later won the game, and the player
of the match was given to the player from the winning team who scored the
highest runs in that match. How many runs were given by extras in total in
that match?
\end{tcolorbox}

\paragraph{Flow.}
\begin{enumerate}[leftmargin=1.7em,itemsep=1pt,topsep=2pt]
    \item \textbf{Candidate formation.} The \YangLiu{main agent} interprets the
    no-bowler-credit clue as an unusual dismissal and narrows the period to
    five candidate international matches involving either
    \emph{handled the ball} or \emph{obstructing the field}.
    \item \textbf{Parallel verification.} It dispatches all five candidates
    concurrently. Every \YangLiu{subagent} receives the same checklist: recover the
    scorecard, count ducks and run-outs, verify the toss and result, verify
    the player of the match, and total the extras.
    \item \textbf{Central synthesis.} The \YangLiu{subagent reports} eliminate four candidates.
    The \YangLiu{main agent} independently revisits the surviving scorecard and the
    unusual-dismissal record, then adds the extras from the two innings.
\end{enumerate}

\paragraph{Parallel candidate decomposition.}
The \YangLiu{main agent} keeps the global constraint set in its own context but
externalizes the expensive scorecard retrieval. Its single dispatch creates
five bounded and mutually exclusive research branches:

\begin{tcolorbox}[
  colback=blue!2,
  colframe=blue!30!black,
  colbacktitle=blue!8,
  coltitle=black,
  title={\YangLiu{Main agent} dispatch (abridged)},
  fonttitle=\bfseries,
  boxrule=0.45pt,
  arc=1mm,
  left=1.5mm,
  right=1.5mm,
  top=1mm,
  bottom=1mm,
  before upper={\small}
]
\textbf{Candidate branches:} Mohsin Khan, Pakistan--Australia Test (1982);
Desmond Haynes, West Indies--India Test (1983); Mohinder Amarnath,
India--Australia ODI (1986); Rameez Raja, Pakistan--England ODI (1987);
and Mohinder Amarnath, India--Sri Lanka ODI (1989).

\medskip
\textbf{Shared verification checklist:} retrieve the full scorecard; count
ducks; count run-outs; identify the toss and match winners; identify the
player of the match; and compute total extras across all innings.
\end{tcolorbox}

This decomposition avoids giving a \YangLiu{subagent} the original open-ended riddle.
Each \YangLiu{subagent} instead receives a named match and a fixed evidence schema,
which makes the returned \YangLiu{subagent reports} directly comparable and cheap for the
\YangLiu{main agent} to integrate.

\paragraph{Evidence returned across the delegation boundary.}
\YangLiu{The subagent assigned to the 1987 Pakistan--England match returns the
following compact findings from its retrieved scorecard and dismissal record.
The box reproduces what crossed the delegation boundary as a qualitative
trajectory excerpt, rather than as independently annotated benchmark evidence.}

\begin{tcolorbox}[
  colback=green!2,
  colframe=green!35!black,
  colbacktitle=green!8,
  coltitle=black,
  title={\YangLiu{Subagent report}: Pakistan vs. England, Karachi, 20 November 1987},
  fonttitle=\bfseries,
  boxrule=0.45pt,
  arc=1mm,
  left=1.5mm,
  right=1.5mm,
  top=1mm,
  bottom=1mm,
  before upper={\small}
]
\begin{itemize}[leftmargin=1.5em,itemsep=1pt,topsep=1pt,parsep=0pt]
    \item Rameez Raja was dismissed for \emph{obstructing the field}, an
    unusual dismissal for which no bowler receives credit.
    \item The three ducks were Phil DeFreitas, Abdul Qadir, and Asif
    Mujtaba; the three run-outs were Mike Gatting, Shoaib Mohammad, and
    Manzoor Elahi.
    \item England won both the toss and the match. Graham Gooch was player
    of the match after scoring 142, the highest score of the match.
    \item The England innings contained 20 extras and the Pakistan innings
    contained 14 extras.
\end{itemize}
\end{tcolorbox}

\YangLiu{The trajectory's parallel subagent reports summarize candidate
elimination as follows. These values are report contents rather than
independently annotated ground truth:}

\begin{center}
\small
\resizebox{\textwidth}{!}{%
\begin{tabular}{lcccccl}
\toprule
Candidate match & Ducks & Run-outs & Toss winner won & PoM condition & Extras & Decision \\
\midrule
Pakistan--Australia, 1982 & 5 & 3 & No  & No  & 65 & Reject \\
West Indies--India, 1983  & 3 & 2 & Draw & --  & 87 & Reject \\
India--Australia, 1986    & 1 & 5 & Yes & No  & 21 & Reject \\
\textbf{Pakistan--England, 1987} & \textbf{3} & \textbf{3} & \textbf{Yes} & \textbf{Yes} & \textbf{34} & \textbf{Retain} \\
India--Sri Lanka, 1989    & 1 & 3 & Yes & Yes & 25 & Reject \\
\bottomrule
\end{tabular}}
\end{center}

\paragraph{\YangLiu{Main agent} synthesis.}
\YangLiu{After receiving the subagent reports, the main agent independently
rechecks the surviving scorecard for the three ducks, three run-outs, toss,
result, and player-of-the-match fields, and separately verifies the dismissal
rule. It then performs the remaining computation, $20+14=34$.}

\begin{tcolorbox}[
  colback=orange!3,
  colframe=orange!45!black,
  colbacktitle=orange!10,
  coltitle=black,
  title={Final answer},
  fonttitle=\bfseries,
  boxrule=0.45pt,
  arc=1mm,
  left=1.5mm,
  right=1.5mm,
  top=1mm,
  bottom=1mm,
  before upper={\small}
]
Only Pakistan vs. England, 2nd ODI at Karachi on 20 November 1987,
satisfies all constraints. England's innings contained 20 extras and
Pakistan's contained 14, giving a combined total of
\textbf{34 runs in extras}.
\end{tcolorbox}

This trace illustrates delegation as active context management rather
than indiscriminate search expansion. The \YangLiu{main agent} preserves the compact
global hypothesis set and final decision, while \YangLiu{subagents} absorb five
independent evidence-retrieval traces and return only constraint-aligned
summaries. Because the branches are independent, their scorecard searches
can be executed in parallel; because they share a reporting schema, the
\YangLiu{main agent} can eliminate candidates without importing their raw browsing
histories into its context.

\end{document}